\documentclass{elsart}
\usepackage{amssymb}
\usepackage[dvips]{epsfig}

\begin{document}

\begin{frontmatter}

\title{Anderson localization: 2-D system in an external magnetic field}

\author{V.N.~Kuzovkov}

\address{Institute of Solid State Physics, University of
Latvia, \\ 8 Kengaraga Street, LV -- 1063 RIGA, Latvia}
\ead{kuzovkov@latnet.lv}

\date{Received \today}

\begin{abstract}
The analytical approach developed by us for the calculation of the
phase diagram for the Anderson localization via disorder [J.Phys.:
Condens. Matter \textbf{14}, 13777 (2002)] is generalized here to
the case of a strong magnetic field when $q$ subbands ($q=1,2,3$)
arise. It is shown that in a line with the generally accepted point
of view, each subband is characterized by a critical point with a
divergent localization length $\xi$ which reveals anomaly in energy
and disorder parameters. These critical points belong to the phase
coexistence area which cannot be interpreted by means of numerical
investigations. The reason for this is a logical incompleteness of
the algorithm used for analysis of a computer modelling for finite
systems in the parameter range where the finite-size scaling is no
longer valid.
\end{abstract}


\end{frontmatter}


\section{Introduction}

In the series of our recent papers
\cite{Kuzovkov02,Kuzovkov04,Kuzovkov06,Kuzovkov07,Reply} an exact
analytic solution for the Anderson localization \cite{Anderson58}
was presented. As it has been commented
\cite{Comment,Suslov05,Suslov06c}, most of our results contradict
the generally-accepted rules of the scaling theory and numerical
modelling. Two particular results were \textit{highly unexpected} in
the Anderson community: (i) \textit{existence} of the Anderson
transition in two dimensional (2-D) disordered systems of
noninteracting electrons, and (ii) the Anderson transition in N-D
($N \geq 2$) dimensions is  of \textit{first-order}, and localized
and conducting states could co-exist.

The result (i) formally does not contradict experiment
\cite{Abrahams01}, where the \textit{unexpected} presence of a
metallic phase in 2-D was observed \cite{Kravchenko04}. It was noted
\cite{Abrahams01} that experiments of the last decade do not support
the prevailing point of view that there can be no metallic state or
metal-insulating transition in a 2-D system. The physics behind
these observations at present is not understood.

The problem is that the experimental facts present not a strong
argument for many theoreticans who believe that they are able to
calculate disordered systems of \textit{noninteracting electrons}
\cite{Markos06}. This problem was studied in the literature by
various methods: scaling theory of localization, perturbation
theory, numerical modelling, etc. The conclusion has been drawn (but
not proved) that a phase transition in experimental systems is
related most likely to the \textit{interactions} (Coulomb,
spin-orbit) neglected in theory. However, this statement is valid
only for theoretical Hamiltonian-based methods. In particular, the
scaling theory of localization \cite{Abrahams79} is a
phenomenological approach and thus
 could potentially be able to take into account strong
particle interactions. The fact that this theory \cite{Abrahams79}
is unable to describe experimental data indicates its serious
problems. It is also possible that a 2-D metal-insulator transition
is also possible for noninteracting electrons
\cite{Kuzovkov02,Suslov05}; however, existing theoretical methods
are not suited for solving such a complex problem.

It should be also noted here that the above-mentioned consensus that
in 2-D all states are localized should be reconsidered in the light
of the conflicting situation. Indeed, as it was noted in a recent
review article \cite{Markos06} devoted to the numerical
investigations of the Anderson localization, practically \textit{all
numerical results} are in contradiction with the \textit{analytical
predictions} (see also discussion in Ref.\cite{Suslov06b}). Despite
author \cite{Markos06} claims that results of numerical analysis
could be treated as obtained ``from first principles'', one should
remember that any numerical analysis is based on certain analytical
predictions. In particular, the necessity to modify the numerical
algorithm for finite-size scaling was discussed \cite{Suslov05}. The
applicability of finite-size scaling studies has also been
questioned \cite{Kantelhardt,Queiroz}. In other words, there is an
obvious conflict between the results of different analytical and
numerical methods.

This is why instead of the discussing point (i) [existence vs
nonexistence of 2-D phase transition] it is more reasonable to
consider consequences of point (ii) in the light of the
applicability of finite-size scaling. This is important since this
is a key method in both numerical investigations of the Anderson
localization \cite{Markos06} and analytical realization of the
finite-size scaling  \cite{Suslov05}. The concept of a finite-size
scaling is taken from theory of phase transitions
\cite{Nightingale,Derrida,Pichard}, and in fact is based on the
extrapolation of the results for finite-size systems (length $L <
\infty$) for the thermodynamic limit, $L=\infty$. It is clear that
as the result of such an extrapolation, a single limiting value
should be obtained. However, this is true only for the second-order
phase transitions and in fact, the finite-size scaling
\cite{Nightingale,Derrida,Pichard} could be applied \textit{only} to
this particular case, but not to the theory of phase transitions in
general.

This leads usually to the assumption of the existence of the
correlation length which is \textit{divergent} at the critical point
\cite{Markos06}. The method cannot be used for the first-order
transitions with a known \textit{multiplicity of solutions}
\cite{Kuzovkov02,Kuzovkov06,Reply} where for an arbitrary set of
system parameters two phases can coexist (localized and delocalized
in our case). As mentioned in Ref. \cite{Kuzovkov02}, a trivial
example is the phase co-existence of ice and water. Only properties
of the \textit{two} pure homogeneous phases are physically
meaningful, whereas an extrapolation of the results for a
heterophase system has no real meaning. However, when the result of
finite-size scaling do not fit to the second-order transition, the
quick (and wrong) conclusion is often drawn that no phase transition
occurs  at all \cite{Kuzovkov02} neglecting a possibility of the
first-order transition.

The type of the phase transition does not follow from any
first-principles \cite{Baxter} and should not be the matter of
speculations \cite{Suslov06b}. This is defined by the exact solution
(if it exists) or careful analysis of information taking into
account all possible alternatives. From the experimental point of
view, the first-order phase transitions are observed relatively
easily. In particular, the direct electrostatic probing
\cite{Ilani1} and photoluminescence spectroscopy \cite{Shashkin94}
clearly show \textit{co-existence} of localized and metallic regions
associated with 2-D metal-insulating transition. New theories are
put forward to address this issue\cite{Meir00, Spivak03}.

In the present paper, we apply our approach
\cite{Kuzovkov02,Kuzovkov04,Kuzovkov06,Kuzovkov07,Reply} to a 2-D
system in an external magnetic field. The field could be included
into a 2-D Hamiltonian by means of a Peierls factor \cite{Markos06}.
This model is used for numerical analysis of the critical Hall
regime in disordered systems in a strong perpendicular magnetic
field. The prevailing viewpoint for this model is that no
metal-insulator transition can occur here. It is believed also that
there is the critical energy $E_c$ in each Landau band at which
electrons are delocalized \cite{Huckestein95}. In other words, this
is nothing but a generalization of our dispute on phase transitions
\cite{Kuzovkov02,Kuzovkov04,Kuzovkov06,Kuzovkov07,Reply} for the
case of an external magnetic field. We demonstrate below that the
model can be exactly solved using the analytical approach
\cite{Kuzovkov02,Kuzovkov04,Kuzovkov06} for a certain range of
magnitude of the control parameter (magnetic flux). As a result, the
generally-accepted viewpoint on the metal-insulator transitions
should be revised. Moreover, unlike the standard opinion
\cite{Huckestein95} that the delocalization at the critical energy
$E_c$ arises entirely due to a strong magnetic field, we show below
that the existence of phase transitions without magnetic field and
the critical energy $E_c$ in the magnetic field are mutually
related.

Note that under the exact solution in this paper we mean only the
calculation of the phase diagram with the localization length. To
solve this problem, we use  the recursive equation for the Cauchy
problem with fixed initial conditions. There are some limitations of
the analytical theory. An exact solution is only possible for the
conventional Anderson model  with a diagonal disorder, where on-site
potentials  are independently and identically distributed. We do not
calculate here transport and other (energy spectrum, etc)
properties, which corresponds to the problem with fixed boundary
conditions (the Dirichlet problem) and has no exact solution.

The paper is organized as follows. In Section 2 the Schr\"odinger
equation is discussed for noninteracting particles in the presence
of a perpendicular magnetic field. The calculation of the joint
correlators is discussed which permits to extract analytically
information on the Lyapunov exponents and to draw the phase diagram.
Since we extend our approach \cite{Kuzovkov02,Kuzovkov04,Kuzovkov06}
for the case of magnetic field, we discuss mostly the necessary
modifications of our approach. This is why knowledge of our previous
papers is prerequisite here. The Lyapunov exponents are found using
theory of functions of complex variables (via search of poles of the
$H(z)$ function on the complex plane). Two conformal mappings are
discussed in Section 3 which permit us to reduce the pole search to
a relatively simple algebraic problem. In Section 4 the main results
are summarized and nontrivial aspects of our theory analyzed.

\section{Main definitions}

\subsection{The model}

We consider a single-band disordered Anderson tight-binding model
with Schr\"o\-din\-ger equation \cite{Markos06}
\begin{eqnarray} \label{Sch}
\psi_{l+1,m}+\psi_{l-1,m}+\exp(-ik_0l)\psi_{l,m+1}+
\exp(ik_0l)\psi_{l,m-1}=(E-\varepsilon_{l,m})\psi_{l,m}
\end{eqnarray}
describing the properties of noninteracting particles in the
presence of a perpendicular magnetic field. The magnetic field
enters the transfer terms connecting nearest neighbors via the
phases, where $k_0=B (ea)^2/\hbar$ is the magnetic flux through the
elementary plaquette of the size $a^2$ (the lattice constant $a$ and
the hopping matrix element are equal to unity). The on-site
potentials $ \varepsilon_{l,m} $ are independently and identically
distributed with existing first two moments, $\left\langle
\varepsilon _{l,m}\right\rangle =0$ and $\left\langle \varepsilon
_{l,m}\varepsilon _{l^{\prime},l}\right\rangle =\sigma ^2\delta_{l,
l^{\prime}}\delta_{m, l}$, where the parameter $\sigma$
characterizes the disorder level.

All moments of the disorder distribution define some physical
properties (energy spectrum, etc), however, phase diagram under
study is defined only be the second moment, this is an exact result
\cite{Kuzovkov02,Kuzovkov04,Kuzovkov06}.  In this paper, along with
the phase diagram, we determine also the Lyapunov exponent $\gamma$,
which is the inverse of the localization length. The Lyapunov
exponent is a typical \textit{order parameter}. It is well known in
the theory of phase transitions that there is no unambiguous
definition of an order parameter. This also holds in our case. Many
different definitions are possible
\cite{Kuzovkov02,Kuzovkov04,Kuzovkov06}, also dependent on all
moments. We use here the $\psi^2$ -definition
\cite{Kuzovkov02,Kuzovkov04,Kuzovkov06}, where the Lyapunov exponent
depends only on the second moment. This is why calculation of the
two first moments of the disorder distribution is sufficient for
solving our problem.

Without disorder ($\varepsilon_{l,m}=0$), the energy spectrum can be
obtained analytically for $k_0=2\pi p/q$, where $p$ and $q$ are
coprime integers (rational number of flux quanta per plaquette). In
system with an external magnetic field each energy band will split
into several Landau subbands ($q$ subbands). Unlike 3-D case, these
subbands are not overlapped with each other \cite{Wang99}.

In a study of disordered system ($\varepsilon_{l,m}\neq 0$) we
discuss a general method, whereas in calculations restrict ourself
to the cases $q=1,2,3$, assuming $p=1$. For $q=1$ $\exp(ik_0l)=1$,
and the relevant Schr\"odinger equation coincides with that without
the magnetic field (only one Landau subband). That is, the problem
is reduced to the results \cite{Kuzovkov02} indicating the existence
of a 2-D metal-insulator transition. The peculiarity of the case of
two Landau subbands ($q=2$) is a lack of the subband gap in the
energy spectrum of the ordered system. For $q=3$ three subbands are
separated by finite width gaps. There is no qualitative difference
between $q=3$ and larger $q$ values.

The peculiarities of the model (\ref{Sch}) caused by the magnetic
field are: (a) the equation contains in general case complex
coefficients and one has to look for complex solutions, (b) the two
directions in a 2-D system (characterized by indeces $l$ and $m$)
formally enter Eq. (\ref{Sch}) asymmetrically which could be
interpreted as anisotropy, (3) the coefficients $\exp(\pm ik_0l)$
entering the Schr\"odinger equation are periodic functions. This is
why the magnetic field problem solution requires a generalization of
the mathematical formalism
\cite{Kuzovkov02,Kuzovkov04,Kuzovkov06,Kuzovkov07}, to be discussed
below.

\subsection{Signals and filter function}\label{Post}

In our previous papers \cite{Kuzovkov02,Kuzovkov04,Kuzovkov06} the
Anderson localization was considered as the particular case of the
generalized diffusion. This process, as any diffusion, is
characterized by the divergence of the average over an ensemble. For
instance, the localization effects lead to the divergence of the
second momentum (diagonal correlators), $\left\langle |\psi_{n,m}|^2
\right\rangle $, as the function of the index $n$ (\textit{discrete
time}),  $n \rightarrow \infty$. The divergence is characterized by
the Lyapunov exponent $\gamma$, which is the inverse of the
localization length,  $\xi=1/\gamma$. Hereafter the average
$\left\langle  \dots \right\rangle$ means the ensemble average over
random potential realizations.

An existence of the so-called fundamental mode responsible for the
correlator $\left\langle  |\psi_{n,m}|^2 \right\rangle $ divergence
was strictly proved in \cite{Kuzovkov06}. It was shown that the
Anderson localization problem could be reduced in the general case
to the so-called \textit{signal theory} characterizing the
\textit{filter function} $h_n$ (localization operator). The
complementary parametrs to $h_n$ are \textit{input signals}
$s^{(0)}_n$ and \textit{output signals} $s_n$
\cite{Kuzovkov02,Kuzovkov04,Kuzovkov06}. These quantities fulfil the
fundamental equation
\begin{equation}\label{eq3}
s_n=\sum_{l=0}^{n}h_{n-l}s^{(0)}_l .
\end{equation}

The input signal characterizes the ideal system with a zero disorder
parameter $\sigma$. Its basic properties are: it is bounded,
$|s^{(0)}_n|< \infty$, as $n \rightarrow \infty$, for physical
(delocalized) band solutions, and is divergent for the formal
(non-physical) solutions. For physical $s^{(0)}_n$  solutions, Eq.
(\ref{eq3}) describes the transformation of the input signal
$s^{(0)}_n$  into another signal $s_n$, called output. The latter
characterizes the disordered system with $\sigma \neq 0$. If for a
given energy $E$ and disorder parameter $\sigma$ the output signal
is bounded, $|s_n|< \infty$, as $n \rightarrow \infty$, this
indicates transformation of input physical solutions into
delocalized ones. In the opposite case, $|s_n| \rightarrow \infty$,
as $n \rightarrow \infty$, the output corresponds to the localized
states.

It was shown \cite{Kuzovkov02,Kuzovkov04,Kuzovkov06} that the
divergence of the output signal with a nonzero disorder is
independent of the input signal, but is a fundamental feature of the
localization operator $h_n$ (the filter function or system
function). As $n \rightarrow \infty$, $|h_n|< \infty$ for
delocalized states, but it is divergent for localized solutions,
$|h_n| \rightarrow \infty$, $n \rightarrow \infty$.

Use of the Z-transform
\begin{equation}\label{H_z}
H(z)=\sum_{n=0}^{\infty}\frac{h_n}{z^n}
\end{equation}
(similar for both input and output signals) permits to reduce the
problem to a search of the poles of the function $H(z)$ of the
complex argument $z$:
\begin{equation}\label{eq4}
H^{-1}(z) = 0 .
\end{equation}
The modified $S^{(0)}(z)$, $S(z)$ and $H(z)$ remain to be called the
input and output signals, and the filter function, respectively,
whereas Eq. (\ref{eq3}) reduces
\begin{equation}\label{eq5}
S(z)=H(z)S^{(0)}(z) .
\end{equation}

It was shown \cite{Kuzovkov02,Kuzovkov04,Kuzovkov06}  that knowledge
of the fundamental characteristics of the system - the filter $H(z)$
- permits to determine the phase diagram of the system (regions of
localized and delocalized solutions), as well as the localization
length. On the contrary, the input and output signals by themselves
are not of great interest. However, the filter, input and output
signals are complementary characteristics: to find the filter
$H(z)$, one has to solve analytically exact equations for signals.
This is possible under certain conditions, but the solution is very
complicated \cite{Kuzovkov06}. However, the calculations could be
considerably simplified after a careful analysis of the structure of
the solution\cite{Kuzovkov06}.

\subsection{The block structure of the Schr\"o\-din\-ger equation}

Let us assume periodicity of the coefficients
\begin{eqnarray}
\exp(ik_0q)=1 ,
\end{eqnarray}
and introduce two integer indices: $n=0,1,...$ (block number or
\textit{discrete time}) and $j=0,1,...,q-1$ (coordinate in the
block), provided $l=nq+j$.

Let us define new amplitudes as
\begin{eqnarray}
\varphi^{(j)}_{n,m}=\psi_{nq+j,m} ,
\end{eqnarray}
and the operators $\mathcal{L}^{(j)}$ which acts on the index $m$
\begin{eqnarray}
\mathcal{L}^{(j)}\varphi^{(j)}_{n,m}=E\varphi^{(j)}_{n,m}-
\sum_{m^{\prime}=\pm 1}
e^{-ik_0jm^{\prime}}\varphi^{(j)}_{n,m+m^{\prime}} .
\end{eqnarray}
Due to periodicity, $\mathcal{L}^{(-1)}=\mathcal{L}^{(q-1)}$ .

Since the coefficients in Eq.(\ref{Sch}) are complex variables, in
order to calculate correlators, we introduce the complex conjugated
functions $\overline{\varphi}^{(j)}_{n,m}$ and operators
$\overline{\mathcal{L}}^{(j)}$,
\begin{eqnarray}
\overline{\mathcal{L}}^{(j)}\overline{\varphi}^{(j)}_{n,l}=E\overline{\varphi}^{(j)}_{n,l}-
\sum_{l^{\prime}=\pm
1}e^{ik_0jl^{\prime}}\overline{\varphi}^{(j)}_{n,l+l^{\prime}} .
\end{eqnarray}

Let us rewrite Eq. (\ref{Sch}) in a recursion form. Inside block
($j=1,2,\dots,q-2$) one gets
\begin{eqnarray}\label{recursM}
\varphi^{(j+1)}_{n,m}=-\varepsilon^{(j)}_{n,m}\varphi^{(j)}_{n,m}+
\mathcal{L}^{(j)}\varphi^{(j)}_{n,m}-\varphi^{(j-1)}_{n,m} ,
\end{eqnarray}
whereas on its boundaries ($j=0$ and $j=q-1$)
\begin{eqnarray}\label{recursM2}
\varphi^{(1)}_{n+1,m}=-\varepsilon^{(0)}_{n+1,m}\varphi^{(0)}_{n+1,m}+
\mathcal{L}^{(0)}_{m,m^{\prime}}\varphi^{(0)}_{n+1,m^{\prime}}-\varphi^{(q-1)}_{n,m}
 ,\\
\varphi^{(0)}_{n+1,m}=-\varepsilon^{(q-1)}_{n,m}\varphi^{(q-1)}_{n,m}+
\mathcal{L}^{(q-1)}_{m,m^{\prime}}\varphi^{(q-1)}_{n,m^{\prime}}-\varphi^{(q-2)}_{n,m}
.
\end{eqnarray}
Similar relations are found for complex-conjugated functions. In
particular, an analog of Eq. (\ref{recursM}) reads
\begin{eqnarray}\label{recursM_1}
\overline{\varphi}^{(j+1)}_{n,l}=-\varepsilon^{(j)}_{n,l}\overline{\varphi}^{(j)}_{n,l}+
\overline{\mathcal{L}}^{(j)}\overline{\varphi}^{(j)}_{n,l}-\overline{\varphi}^{(j-1)}_{n,l}
.
\end{eqnarray}

\subsection{Correlators}

Following our approach \cite{Kuzovkov06}, the diagonal correlators
$\left\langle|\varphi^{(j)}_{n,m}|^2 \right\rangle$ should be
calculated. To this end, a complete set of equations for the two
types of correlators has to be analyzed:
\begin{eqnarray}
x^{(j)}(n)_{m,l}=\left\langle \varphi^{(j)}_{n,m}
\overline{\varphi}^{(j)}_{n,l}\right\rangle ,\\
y^{(j)}(n)_{m,l}=\left\langle \varphi^{(j)}_{n,m}
\overline{\varphi}^{(j-1)}_{n,l}\right\rangle .
\end{eqnarray}
Here $\overline{\varphi}^{(-1)}_{n,l} \equiv
\overline{\varphi}^{(q-1)}_{n-1,l}$,
$\overline{x}^{(j)}(n)_{m,l}=x^{(j)}(n)_{l,m}$.

The equations for correlators can be derived relatively easily. Say,
to define  $x^{(j)}(n)_{m,l}$  one has to multiply the left and
right sides of the above-given equations, in particular,
Eqs.(\ref{recursM}) and (\ref{recursM_1}) are used inside the same
block. Then, the obtained relations should be averaged over the
ensemble of random potentials:
\begin{eqnarray}
x^{(j+1)}(n)_{m,l}=\left\langle \varepsilon^{(j)}_{n,m}
\varepsilon^{(j)}_{n,l}\right\rangle \left\langle
\varphi^{(j)}_{n,m}\overline{\varphi}^{(j)}_{n,l}\right\rangle + \\
\nonumber +\left\langle
\mathcal{L}^{(j)}\varphi^{(j)}_{n,m}\cdot\overline{\mathcal{L}}^{(j)}
\overline{\varphi}^{(j)}_{n,l}\right\rangle+ \left\langle
\varphi^{(j-1)}_{n,m}\overline{\varphi}^{(j-1)}_{n,l}\right\rangle -
\\ \nonumber
-2\left\langle\mathcal{L}^{(j)}\varphi^{(j)}_{n,m}\cdot\overline{\varphi}^{(j-1)}_{n,l}\right\rangle
-2\left\langle\varphi^{(j-1)}_{n,m}\cdot\overline{\mathcal{L}}^{(j)}
\overline{\varphi}^{(j)}_{n,l}\right\rangle -\\ \nonumber
-2\left\langle\varepsilon^{(j)}_{n,m}\right\rangle\left\langle
\varphi^{(j)}_{n,m}\cdot[\overline{\mathcal{L}}^{(j)}\overline{\varphi}^{(j)}_{n,l}-
\overline{\varphi}^{(j-1)}_{n,l}]\right\rangle  -\\ \nonumber
-2\left\langle\varepsilon^{(j)}_{n,l}\right\rangle\left\langle
\overline{\varphi}^{(j)}_{n,l}\cdot[\mathcal{L}^{(j)}\varphi^{(j)}_{n,m}-\varphi^{(j-1)}_{n,m}]\right\rangle
.
\end{eqnarray}
When doing so, the \textit{causality principle}
\cite{Kuzovkov02,Kuzovkov04,Kuzovkov06} is taken into account, which
means that all amplitudes $\varphi^{(j)}_{n,m}$,
$\overline{\varphi}^{(j)}_{n,l}$  on the r.h.s. of obtained equation
are statistically independent with respect to the potentials
$\varepsilon^{(j)}_{n,m}$,  $\varepsilon^{(j)}_{n,l}$. In other
words, when calculating the average quantities on the r.h.s. of
equations, only amplitude correlations are essential. The
calculation of averages over the potentials is quite trivial: we use
know averages for the first two moments, $\left\langle
\varepsilon^{(j)}_{n,m}\right\rangle=0$ and
\begin{equation}\label{eps2}
\left\langle
\varepsilon^{(j)}_{n,m}\varepsilon^{(j^{\prime})}_{n^{\prime},m^{\prime}}\right\rangle=\sigma^2
\delta_{j,j^{\prime}}\delta_{n,n^{\prime}}\delta_{m,m^{\prime}} ,
\end{equation}
provided  $\sigma$ remains the disorder parameter.

Proceeding this way, one gets on the r.h.s. of the derived equations
along with the correlators $x^{(j)}(n)_{m,l}$ sought for, also
additional terms with the correlators $y^{(j)}(n)_{m,l}$. The
equations for $y^{(j)}(n)_{m,l}$ could be derived in a similar way,
r.h.s. are expressed through the correlators $x^{(j)}(n)_{m,l}$ and
$y^{(j)}(n)_{m,l}$. That is, the equation set is complete.

Note that the boundary conditions for the amplitude in the general
form, $\psi_{0,m}=0$ è $\psi_{1,m}=\alpha_m$  are reduced to the
boundary conditions for the correlators, e.g. $x^{(0)}(0)_{m,l}=0$,
$x^{(1)}(0)_{m,l}=\alpha_m \overline{\alpha}_l$.  Due to an
arbitrary choice of the field $\alpha_m$, the \textit{translational
invariance} does not occur. However, the complete analytical
solution of the correlator equations is possible by means of the
\textit{double} Fourier transform and Z-transform \cite{Kuzovkov06}.
The same is true for the problem under study. This approach is exact
but rather lengthy.

The derivation could be, however, considerably reduced, taking into
account the important result of our paper \cite{Kuzovkov06}. In
fact, we introduced the correlators as the tool in our study, but
all we need is only the term in the diagonal correlators
$x^{(j)}(n)_{m,m}$ called the \textit{fundamental mode}; it is
divergent for the delocalized states. Namely this quantity
corresponds to signals $s_n$ in Eq. (\ref{eq3}).

The fundamental mode is \textit{invariant} with respect to the
argument shift in the boundary condition  $\alpha^{\prime}_{m}\equiv
\alpha_{m+m_0}$. The equations for signals could be easily obtained
by the formal replacement of the boundary conditions in equations
for the correlators \cite{Kuzovkov06}: $x^{(1)}(0)_{m,l}=\alpha_m
\overline{\alpha}_l$ is replaced for
$x^{(1)}(0)_{m,l}=\underline{\alpha_m
\overline{\alpha}_l}=\Gamma_{m-l}$. Here the function  $\Gamma_r$
corresponds to the Fourier transform $\Gamma(k)=|\alpha(k)|^2$.

After such a replacement the problem becomes translationally
invariant, the correlators depend only on the argument divergence
$r=m-l$
\begin{eqnarray}
x^{(j)}(n)_{m,l} \rightarrow x^{(j)}(n)_r , \\
y^{(j)}(n)_{m,l} \rightarrow y^{(j)}(n)_r ,
\end{eqnarray}
whereas the diagonal correlators become independent on the $m$-argument
(fundamental mode) and transform into a set of signals
\begin{equation}\label{signals}
x^{(j)}(n)_{m,m}\rightarrow s(j,n) .
\end{equation}
For simplicity we use hereafter the correlator symmetrization which
does not change the diagonal elements sought for
\begin{eqnarray}
\hat{x}^{(j)}(n)_r=\frac{1}{2}[x^{(j)}(n)_r+x^{(j)}(n)_{-r}] ,\\
\hat{y}^{(j)}(n)_r=\frac{1}{2}[y^{(j)}(n)_r+\overline{y}^{(j)}(n)_{-r}]
.
\end{eqnarray}
By definition,
\begin{equation}\label{cor-sig}
\hat{x}^{(j)}(n)_0=s(j,n) .
\end{equation}

\subsection{The Z-transform and the Fourier transform}

The analytical solution is based on the use of the two type of
algebraic transforms: the Z-transform
\begin{eqnarray}
S(j,z)=\sum_{n=0}^{\infty} \frac{s(j,n)}{z^{nq+j}} ,\\
\hat{x}^{(j)}(z)_r =\sum_{n=0}^{\infty}
\frac{\hat{x}^{(j)}(n)_r}{z^{nq+j}} ,
\end{eqnarray}
and the Fourier transform,
\begin{eqnarray}
\hat{x}^{(j)}(z)_r =\int_{-\pi}^{\pi}
\frac{dk}{2\pi}\hat{X}^{(j)}(z,k)e^{-ikr} .
\end{eqnarray}

The Eq. (\ref{cor-sig})  for the correlator-signal relation
transforms into
\begin{equation}\label{cor-sig2}
\hat{x}^{(j)}(z)_0= S(j,z) ,
\end{equation}
which gives also the following useful integral relation:
\begin{eqnarray}\label{Inegris}
\int_{-\pi}^{\pi} \frac{dk}{2\pi}\hat{X}^{(j)}(z,k) = S(j,z) .
\end{eqnarray}

\subsection{Equations}

As a result, we obtain the equation set for the correlators
($j=0,1,\dots,q-1$)
\begin{eqnarray}\label{hatX}
z\hat{X}^{(j+1)}(z,k)-\Gamma(k)\delta_{j,1}= \sigma^2 S(j,z)+
 \\ \nonumber +
z^{-1}\hat{X}^{(j-1)}(z,k)+ \mathcal{L}^2_j(k)\hat{X}^{(j)}(z,k)
-2 \mathcal{L}_j(k)\hat{Y}^{(j)}(z,k) ,\\
z\hat{Y}^{(j+1)}(z,k)=-\hat{Y}^{(j)}(z,k)+\mathcal{L}_j(k)
\hat{X}^{(j)}(z,k) .\label{hatY}
\end{eqnarray}
Here the functions
\begin{equation}\label{Lkj}
\mathcal{L}_j(k)=E - 2\cos (k-j k_0)
\end{equation}
arise from the Fourier transform of the operators
$\mathcal{L}^{(j)}$. The periodic conditions take place with respect
to the index $j$: $\hat{X}^{(q)}(z,k)\equiv \hat{X}^{(0)}(z,k)$.

The solution algorithm is quite simple. For a given $q$ a set of
$2q$ equations (\ref{hatX}), (\ref{hatY}) has to be solved. As a
result, one gets $q$ relations for the correlators
$\hat{X}^{(j)}(z,k)$ which are linear in signals $S(j,z)$ and
function $\Gamma(k)$. After use of Eq. (\ref{Inegris}) we obtain the
set of linear equations for $q$ signals $S(j,z)$. The coefficients
here are expressed through integrals. Taking into account
coefficient symmetry with respect to index transformation, all
partial signals $S(j,z)$ could be combined into the \textit{total
signal}
\begin{eqnarray}
S(z)=\sum_{j=0}^{q-1}S(j,z) .
\end{eqnarray}

In all cases the equation for  $S(z)$ is reduced to a general form,
Eq.(\ref{eq5}), and $H(z)=1$ as $\sigma=0$ takes place. As a result,
we derive equations for the filter function $H(z)$  and the input
signal $S^{(0)}(z)$. Since for calculating the phase diagram the
filter function $H(z)$ is sufficient, we do not discuss here the
bulky equations for the input signals $S^{(0)}(z)$. The more so, the
analysis of input signals $S^{(0)}(z)$ \cite{Kuzovkov06} leads to
the trivial conclusions: the input signals are bounded if the wave
functions without disorder form the band, and unbound outside the
band. However, the existence range of the band without disorder
could be found by means of standard methods, without complicated
analysis of the asymptotic behaviour of the signals.

\subsection{Main relations}

Let us consider three cases, $q=1,2,3$. For  $q=1$ one gets
$k_0=2\pi$, $\mathcal{L}_0(k)=\mathcal{L}(k)$, where
\begin{equation}
\mathcal{L}(k) = E - \cos(k) .
\end{equation}
To illustrate the derivation method of the equations for $S(z)$, we
consider a simple case of one Landau subband ($q=1$). For other
cases, only the final relations are presented.

For $q=1$, let us denote $\hat{X}^{(0)}(z,k)=\hat{X}(z,k)$,
$\hat{Y}^{(0)}(z,k)=\hat{Y}(z,k)$, respectively. One gets the
equation set
\begin{eqnarray}
(z-z^{-1})\hat{X}(z,k)-\Gamma(k)= \sigma^2 S(z)+
\mathcal{L}^2(k)\hat{X}(z,k)
-2 \mathcal{L}(k)\hat{Y}(z,k) ,\\
(z+1)\hat{Y}(z,k)=\mathcal{L}(k) \hat{X}(z,k) .
\end{eqnarray}
This set is solved trivially with respect to $\hat{X}(z,k)$:
\begin{eqnarray}
\frac{(z-1)}{(z+1)}\left[\frac{(z+1)^2}{z}-\mathcal{L}^2(k)\right]\hat{X}(z,k)=\Gamma(k)+\sigma^2
S(z) .
\end{eqnarray}
The equation for the signal $S(z)$ could be derived using the
Eq.(\ref{Inegris}):
\begin{eqnarray}\label{Inegris2}
\int_{-\pi}^{\pi} \frac{dk}{2\pi}\hat{X}(z,k) = S(z) .
\end{eqnarray}
The resulting equations were discussed in our Ref.\cite{Kuzovkov02}.

The results for $q=1,2,3$ could be presented in a general form
\begin{eqnarray}
\frac{1}{H(z)} = 1-\sigma^2
\frac{(z^q+1)}{(z^q-1)}\int_{-\pi}^{\pi} \frac{d k}{2\pi}
\frac{\left [N_1(E,z)-N_2(E,z)\mathcal{L}(k)\right
]}{(z^q+1)^2/z^q-\mathcal{L}^2(k)} . \label{Hq_all}
\end{eqnarray}
The coefficient $N_1(E,z)$, $N_2(E,z)$ here are defined below.

For $q=2,3$ $\mathcal{L}(k)$ reads:
\begin{eqnarray}
\mathcal{L}(k)=\mathcal{L}_0(k)\mathcal{L}_1(k)-2 ,
\end{eqnarray}
\begin{eqnarray}
\mathcal{L}(k)=\mathcal{L}_0(k)\mathcal{L}_1(k)\mathcal{L}_3(k)-
[\mathcal{L}_0(k)+\mathcal{L}_1(k)+\mathcal{L}_3(k)] .
\end{eqnarray}

Simple trigonometrical transformations show that
\begin{eqnarray}
\mathcal{L}(k)=\delta-2\cos(qk) .
\end{eqnarray}
The parameter $\delta=\delta(E)$  equals $\delta=E$ ($q=1$),
$\delta=E^2-4$ ($q=2$) and $\delta=E(E^2-6)$ ($q=3$). Irrespective
of the $q$ value, the parameter $\delta \in [-4,4]$ for each subband
in an ideal system ($\sigma=0$).

Notice that the integrand in Eq. (\ref{Hq_all}) contain always
$N_1(E,z)-N_2(E,z)\mathcal{L}(k)$ linear in $\mathcal{L}(k)$. This
simplification arises due to our regular use of symmetry and
trigonometrical transformations.

In particular, for $q=2$ the intermediate calculations contain the
following intergrals
\begin{eqnarray}\label{Inegris3}
J=\int_{-\pi}^{\pi}
\frac{dk}{2\pi}\frac{\mathcal{L}^2_j(k)}{(z^2+1)^2/z^2-\mathcal{L}^2(k)}
.
\end{eqnarray}
It could be easily shown that the integral does not depend on the
index $j$. In particular, in the integral numerator, Eq.
(\ref{Inegris3}), $\mathcal{L}^2_j(k)$ could be replaced by
\begin{eqnarray}
\mathcal{L}^2_j(k)\Rightarrow
\frac{1}{2}\sum_{j=0}^1\mathcal{L}^2_j(k) = 2E^2-2- \mathcal{L}(k).
\end{eqnarray}

Similarly, in the case $q=3$, one can use the result invariance with
respect to the cyclic index transpositions:
\begin{eqnarray}
\mathcal{L}_j(k)\mathcal{L}(k)\Rightarrow
\frac{1}{3}\sum_{j=0}^2\mathcal{L}_j(k)\mathcal{L}(k) = E\mathcal{L}(k) , \\
\mathcal{L}^2_j(k)\Rightarrow
\frac{1}{3}\sum_{j=0}^2\mathcal{L}^2_j(k) = 2+E^2 , \\
(\mathcal{L}_1(k)\mathcal{L}_2(k)-1)^2\Rightarrow
6+3E^4-14E^2-2E\mathcal{L}(k) .
\end{eqnarray}

As a result, one obtains the coefficients in the numerator of the
integral, Eq. (\ref{Hq_all}). For $q=1$ one gets $N_1(E,z)=1$,
$N_2(E,z)=0$. For $q=2$
\begin{eqnarray}
N_1(E,z)=\frac{(z^2+1)}{z}+2(E^2-1) , \\
N_2(E,z)=\frac{(z-1)^2}{(z^2+1)} .
\end{eqnarray}
Lastly, in the case $q=3$
\begin{eqnarray}
N_1(E,z)=(6-14E^2+3E^4)+(z^2+z^{-2})+(2+E^2)(z+z^{-1}) , \\
N_2(E,z)=2E\frac{z^2-2z+1}{z^2-z+1} .
\end{eqnarray}

\section{Conformal mapping}

\subsection{The first conformal mapping}

Note that the input and output signals are real values. According to
signal theory \cite{Weiss}, the roots $z_i$ of Eq.(\ref{eq4}) either
lie on the real axis or arise as complex conjugate pairs,
$z_i,\overline{z}_i$. The properties of the filter function $H(z)$
and physical interpretation of the mathematical solution are
determined by the location of the \textit{largest} (per modulus)
root of Eq. (\ref{eq4}), $z_{max}=\max \{|z_i |\}$. As $z_{max} >1$,
the filter $H(z)$ is \textit{unstable} and corresponds to the
localized states, otherwise for $z_{max} \leq 1$ (\textit{stable
filter}) and delocalized  states occur. This is why, to determine
the stability region boundaries of the filter $H(z)$, we restrict
ourselves to the $z$ values in the \textit{upper half-plane}, $0
\leq \arg \,z \leq \pi$.

Search of the complex roots of Eq. (\ref{eq4}) could be simplified
by performing two conformal mappings.

First, we change the complex variable $z$ for the new parameter
$\zeta=z^q$.
\begin{enumerate}
\item With a magnetic field for $k_0=2\pi$ and without magnetic field (
formally, $q=1$) this is a trivial
transformation, $\zeta=z$, so that the $\zeta$  variable is defined
in the upper half-plane.

\item For  $q=2$ we use the relation $\zeta =z^2$, so the $\zeta$
variable is defined on a whole complex plane,  $0 \leq \arg \,\zeta
< 2\pi$.  The inverse transformation gives $z=\zeta^{1/2}$,
respectively.

\item In the case $q=3$ we divide the upper half-plane where $z$ is
defined, into two regions. (a) In the first region, where $0 \leq
\arg \, z < 2\pi/3$  we define $\zeta=z^3$, so that $\zeta$ again is
defined on a whole complex plane  $0 \leq \arg \,\zeta < 2\pi$. The
inverse transformation yields $z=\zeta^{1/3}$. (b) In the second
region, where $2\pi/3 \leq \arg \, z < \pi$  we also assume
$\zeta=z^3$, but now define the inverse transformation as
$z=\zeta^{1/3}\cdot \exp(i 2\pi/3)$.  In this region the variable
$\zeta$ is defined in the upper half-plane.
\end{enumerate}

As a result, we obtain $\zeta$-presentation of the filter function
$H(\zeta)$. Next we seek the roots of the equation $H^{-1}(\zeta)=0$
and obtain for $q=1,2,3$
\begin{eqnarray}
\frac{1}{H(\zeta)} = 1-\sigma^2
\frac{(\zeta+1)}{(\zeta-1)}\int_{-\pi}^{\pi} \frac{d k}{2\pi}
\frac{\left [N_1(E,z)-N_2(E,z)\mathcal{L}(k)\right
]}{(\zeta+1)^2/\zeta-\mathcal{L}^2(k)} .\label{Hq_zeta}
\end{eqnarray}

\subsection{Second conformal mapping}

When calculating intergrals in Eq. (\ref{Hq_zeta}), it is convenient
to use the relation well-known for functions of complex variables
\begin{equation}\label{eq103}
  \int_{-\pi}^{\pi} \frac{dk}{2\pi}\frac{1}{\tau \pm 2\cos(qk)} =
  \frac{1}{i \sqrt{4-\tau^2}} .
\end{equation}
This relation holds for
arbitrary $q=1,2,\dots$ provided $Im \, \tau \geq 0$ (the complex
variable  $\tau$ is defined in the \textit{upper half-plane}) .

The integral in Eq.(\ref{Hq_zeta}) defines the function which is
non-analytic on the unit circle  $|\zeta|=1$
\cite{Kuzovkov02,Kuzovkov04}. When performing the inverse
Z-transform, this leads to a \textit{double solution}. It is
convenient \cite{Kuzovkov02,Kuzovkov04} to perform a second
conformal mapping
\begin{eqnarray}\label{wz}
w=\pm (\zeta^{1/2}+\zeta^{-1/2}) ,
\end{eqnarray}
where the choice of a sign defines one of the two filter functions,
$H_{\pm}(w)$. For $w=+(\zeta^{1/2}+\zeta^{-1/2})$ (or for
$w=-(\zeta^{1/2}+\zeta^{-1/2})$), the outer (inner) part of the unit
circle, $|\zeta|=1$, transforms onto the \textit{upper half-plane}
$w=u+iv$, $v= Im \, w\geq 0$. The circle itselfs maps onto the
interval $u \in [-2,2]$ on the real axis ($v=0$).  The
transformation inverse to Eq. (\ref{wz})  reads \cite{Kuzovkov02}
\begin{eqnarray}\label{zw}
\zeta=-1+\frac{w^2}{2} \pm\frac{w i}{2}\sqrt{4-w^2} .
\end{eqnarray}

\subsection{Parametric representation of the pole diagram}

As shown \cite{Kuzovkov02,Kuzovkov04}, physical information on the
localization could be obtained from the $H_{\pm}(w)$  using standard
methods for the functions of complex variables. The key issue here
is the location of poles on the complex plane (pole diagram)
\cite{Weiss}.

Let us present $H_{\pm}(w)$ in the following form
\begin{eqnarray}\label{R}
H^{-1}_{\pm}(w)=1-\sigma^2 R_{\pm}(w) .
\end{eqnarray}
The main idea is quite simple. The function  $H_{\pm}(w)$  has its
poles where
\begin{eqnarray}\label{R2}
\sigma^2 R_{\pm}(w) =1
\end{eqnarray}
takes place. So, we have the parametric $w$-representation of the
pole diagram. The relevant expressions for $R_{\pm}(w)$  are quite
simple and allow analytical solution. In particular, for $q=2$ one
gets
\begin{eqnarray}
R_{+}(w)=\frac{1}{
\sqrt{w^2-4}}\left[\frac{4+\delta}{\sqrt{(\delta-w)^2-4}}+
\frac{2+\delta+w}{\sqrt{(\delta+w)^2-4}}\right] ,\\
R_{-}(w)=\frac{1}{
\sqrt{4-w^2}}\left[\frac{4+\delta}{\sqrt{4-(\delta+w)^2}}+
\frac{2+\delta-w}{\sqrt{4-(\delta-w)^2}}\right] .
\end{eqnarray}
The diagrammatic technique for seeking roots of Eq. (\ref{R2}) was
described in detail in appendix of Ref.\cite{Kuzovkov04}. This is
why we present below only the main results.

\section{Results}

\subsection{Stable filter}

The function $H_{-}(w)$ (so-called \textit{stable filter}
\cite{Weiss}) defines the existence region of the delocalized
states. General results of its study (Appendix in Ref.
\cite{Kuzovkov04}) can be summarized as follows. The $H_{-}(w)$
allow physical interpretation if the equation $\sigma^2 R_{-}(w) =1$
either has no roots (which is possible only for space dimensions
higher than two), or the roots of the equation, $w=u+iv$, are real
and lie in the interval $w=u \in [-2,2]$.

That is, in our case it is sufficient to find the existence region
of the \textit{real roots} $w$ of the equation $\sigma^2 R_{-}(w)
=1$ in the interval $w=u \in [-2,2]$, provided the absence of other
(complex) roots. As a result, one can determine the function
$\sigma=\sigma_0(\delta)$ (or $\sigma=\sigma_0(E)$, when changing
the $\delta$ variable for $E$) which gives the critical disorder
destroying the localized states. This function is plotted below,
Fig.1, when analyzing the pase diagrams for $q=1,2,3$.

For the illustration we present some results for the case $q=2$ (two
Landau subbands) which were obtained partly analytically, partly
using numerical methods. The equation $\sigma^2 R_{-}(w) =1$  for
$\delta=0$ has two real roots in the interval $w=u \in [-2,2]$
provided the disorder parameter $0 \leq \sigma \leq
\sigma_0(\delta=0)=0.828$. As $\sigma \geq \sigma_0(\delta=0)$,
equation has complex roots having no physical interpretation. As
already was mentioned, the interval  $\delta \in [-4, 4]$
corresponds to the old band (for zero disorder). In the intervals
$-4 \leq \delta \leq \delta_0$ ($\delta_0=-0.741$) and $2 \leq
\delta \leq 4$ the function $\sigma_0(\delta)\equiv 0$, that is
\textit{infinitesimally weak disorder} already destroys all
delocalized states. However, the delocalized states are allowed in
the interval $0 < \delta < 2 $, provided the disorder parameter
$\sigma$  is below some critical value. Note, that the function
$\sigma_0(\delta)$ is multi-value in the region $\delta \in
(\delta_0,0)$  and has a characteristic triangle shape. The
delocalized states are destroyed in the beginning even due to very
weak disorder, as the $\sigma$ parameter grows, formation of the
delocalized state phase is possible. The point $\delta_0$  on the
energy scale ($\delta=E^2-4$) corresponds to the $|E|=E_0=1.805$.
The formation of a phase of delocalized states is possible in the
energy interval $|E|\in (1.805,2.449)$, respectively.

\subsection{Unstable filter}

The localization is characterized by the \textit{unstable filter}
$H_{+}(w)$ \cite{Kuzovkov04}. The main result is quite simple
\cite{Kuzovkov02,Kuzovkov04}:  we are seeking one specific root of
the equation  $\sigma^2 R_{+}(w) =1$.  This root corresponds to the
pole maximally distant from the coordinate origin, $z_0=z_{max}$,
and could be related to the localization length $\xi$ via a simple
equation: $z_0=\exp(2\gamma)$ \cite{Kuzovkov02} (where the Lyapunov
exponent  $\gamma=1/\xi$). After two conformal mappings the point
$z_0$ maps onto $w_0=2\cosh(2q\gamma)$ and the mathematical problem
is reduced to seeking a real root of the equation  $\sigma^2
R_{+}(w) =1$.

It could be shown that the root $w_0$ sought lies on the real axis
in the interval $w=u \geq u_0$ where
\begin{eqnarray}
u_0 = 2+|\delta| .
\end{eqnarray}
The value of $\sigma=0$ corresponds to the  $w_0=u_0$, and $w_0$
monotonically increases, as $\sigma$ increases. That is, in the
limiting case of $\sigma \rightarrow 0$  the Lyapunov exponent
$\gamma=\gamma(\delta,\sigma)$  becomes
\begin{equation}\label{eq112}
  \gamma(\delta,0)=\frac{1}{2q}\cosh^{-1}(1+|\delta|/2) .
\end{equation}
The minimal value of $u_0=2$ occurs for $\delta=0$, that is the
critical energy values  $E_c$ (\textit{one per each subband}) being
defined by the equation  $\delta(E)=0$ are well defined
(\textit{anomaly} discussed below). Here $\gamma(\delta=0,0)=0$,
i.e. the localization length $\xi$ is \textit{divergent} at the
critical points $E_c$, which are equal to: $E_c=0$ ($q=1$), $E_c=\pm
\sqrt{2}$ ($q=2$), and $E_c=0, \pm \sqrt{6}$ ($q=3$).

In the limit of a small  $\sigma$ one gets
\begin{equation}\label{eq114}
  \gamma(\delta=0,\sigma) \propto \sigma .
\end{equation}
From Eq. (\ref{eq112}) for $|\delta|\ll 1 $  one gets
\begin{equation}\label{eq115}
\gamma(\delta,0) \propto \sqrt{|\delta|} ,
\end{equation}

Omitting the proof details, let us discuss the results for the
Lyapunov exponent $\gamma(E,\sigma)$ (i.e. using $E$ instead of
parameter $\delta$) as shown in Fig.2, for $q=1,2,3$.

 Eq. (\ref{eq115}) suggests that in the vicinity of the
critical point $E_c$ (taking into account $|\delta| \propto
|E-E_c|$) in the limit $\sigma \rightarrow 0$  one gets $\gamma(E,0)
\propto |E-E_c|^{1/2}$. That is, we expect a strong anomaly in the
Lyapunov exponent  $\gamma(E,\sigma)$ behaviour. In surface plot,
Fig.2, this anomaly looks like deep canyons around the points
($E=E_c,\sigma=0$). The canyon banks are very steep, therefore
minimal values of the Lyapunov exponents lie on the canyon bottom,
i.e. form the line. This anomaly completely confirms the general
point of view that the localization length diverges at a single
energy point at the center of the Landau band \cite{Huckestein95}.

\subsection{Phase diagram}

As was mentioned, a study of the $H_{-}(w)$-function allows to
determine the existence region for the delocalized states, whereas
$H_{+}(w)$ characterizes the localized states. It was found
\cite{Kuzovkov04}, however, that the localized states exist at all
energies and non-zero disorder parameters.

\begin{figure}[htbp]
  \begin{center}
  \fbox{\epsfig{file=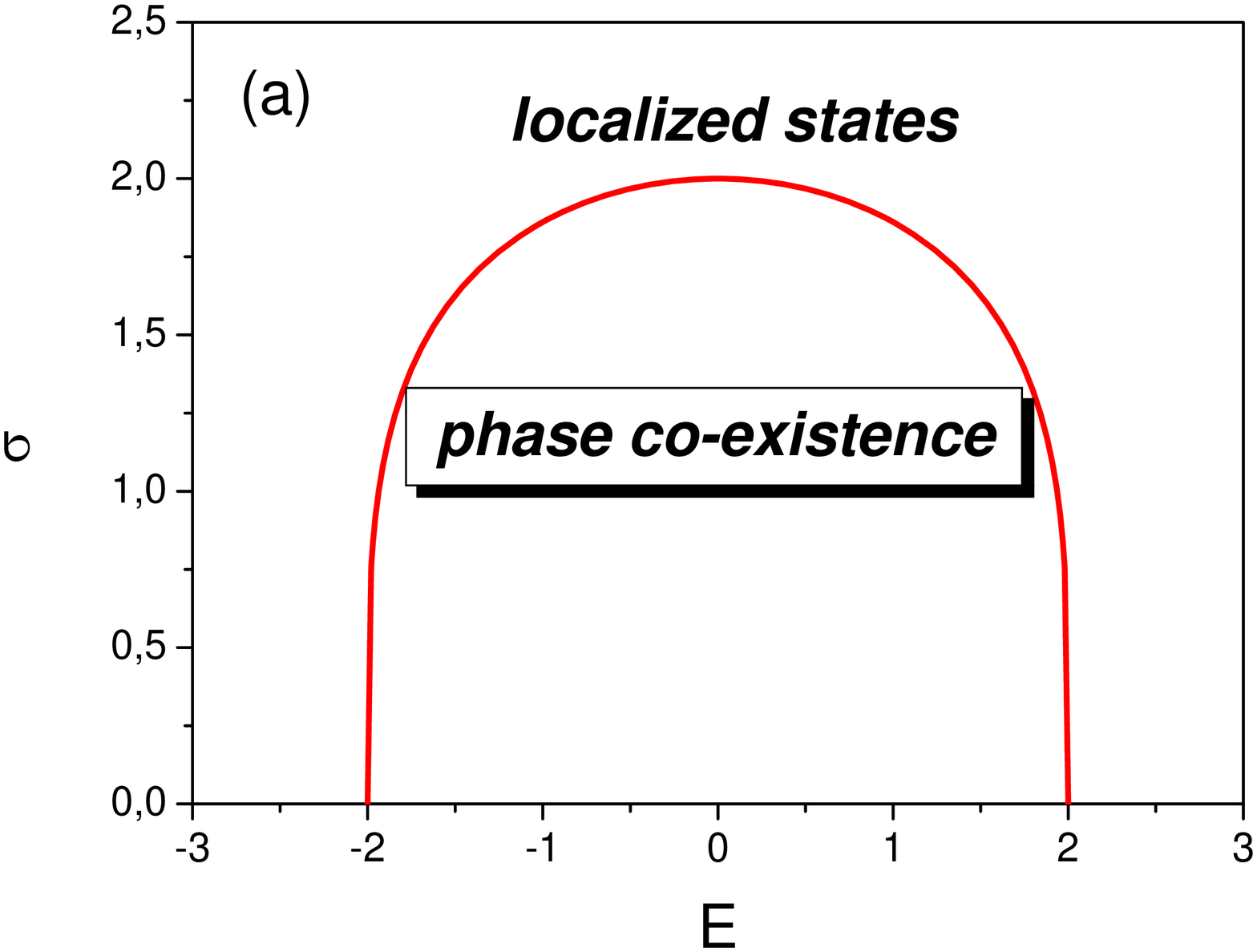,angle=0,width=4cm}}
    \fbox{\epsfig{file=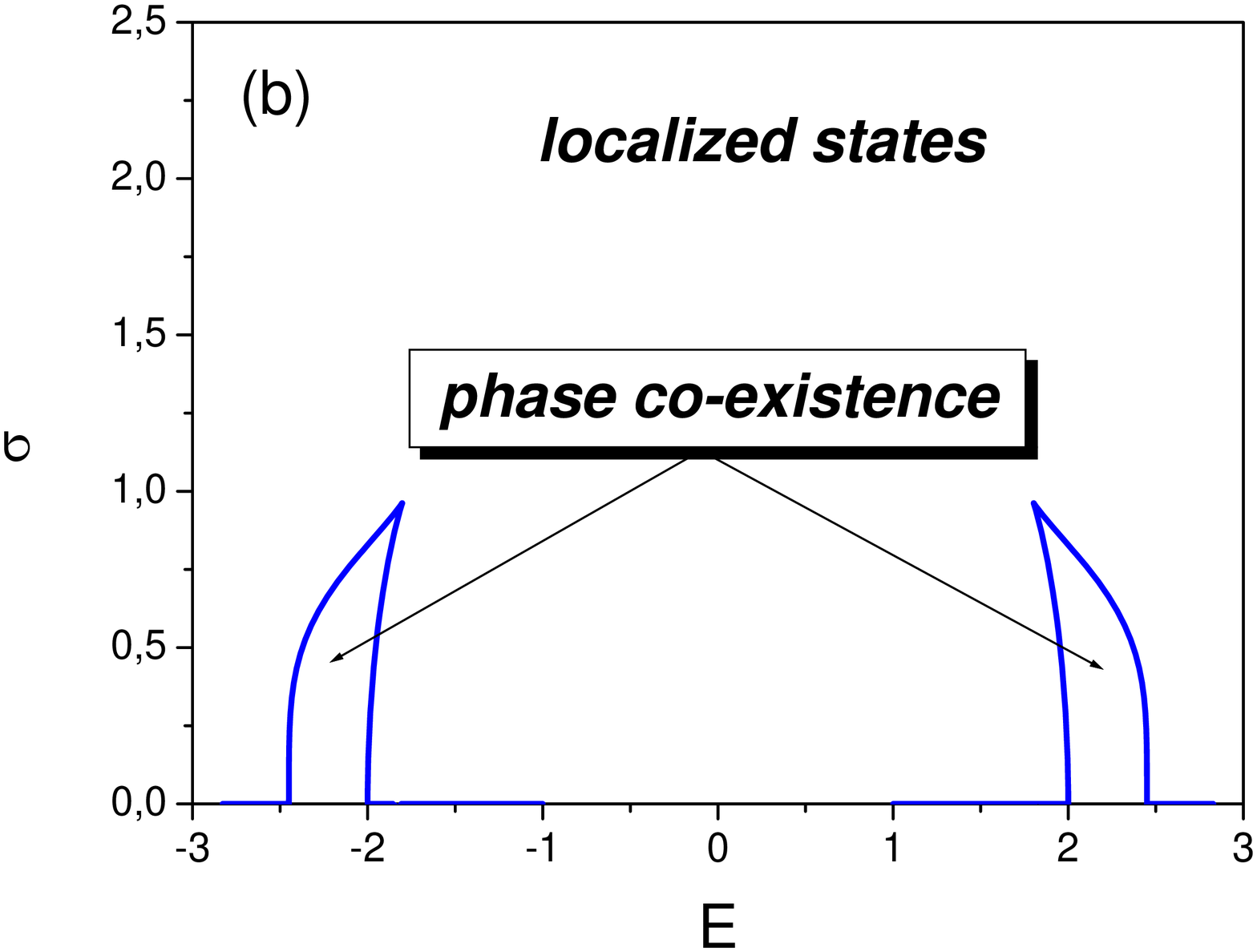,angle=0,width=4cm}}
    \fbox{\epsfig{file=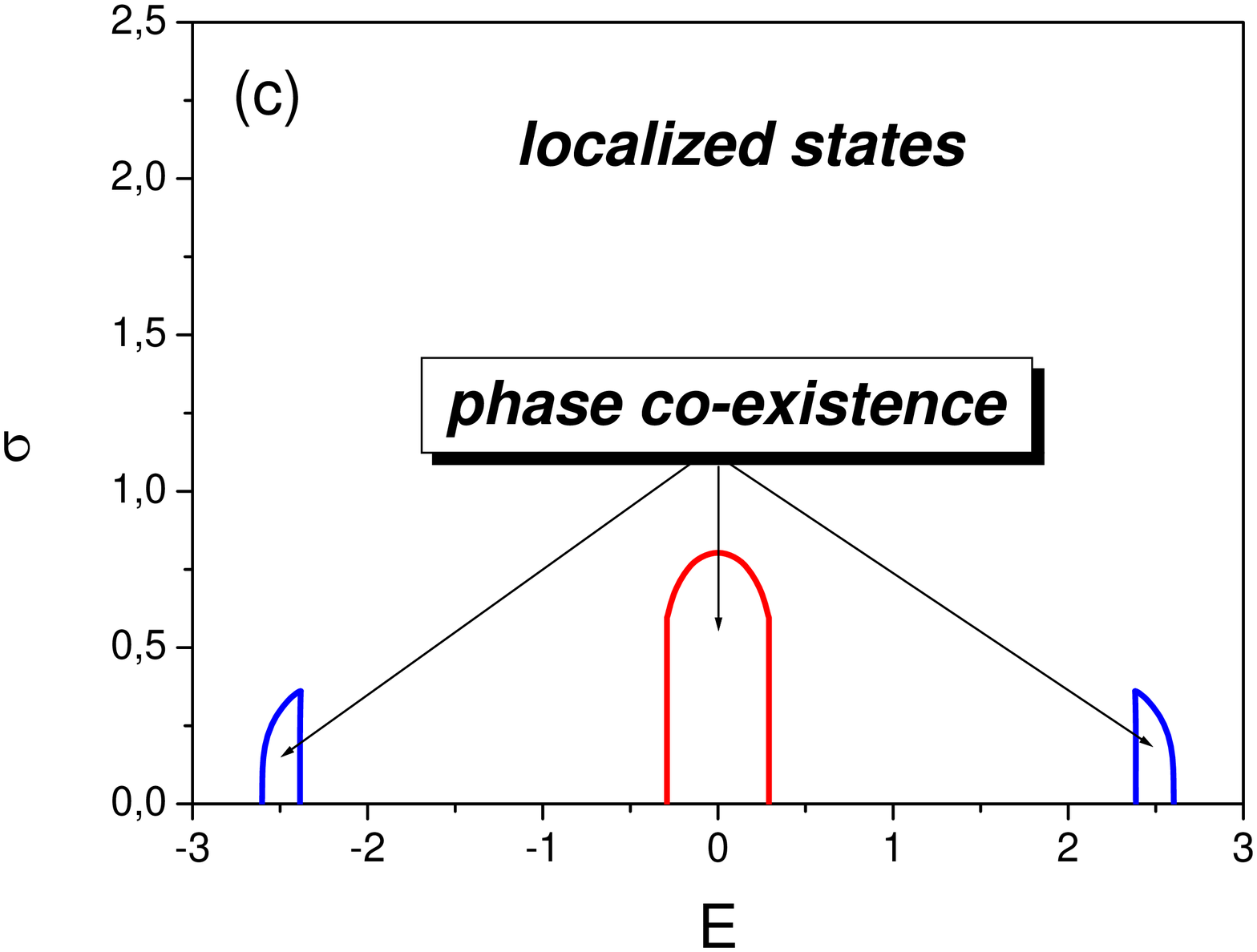,angle=0,width=4cm}}
    \caption{
    Phase diagram for $q=1$ (a), $q=2$ (b) and $q=3$ (c).
      }
    \label{fig: 1}
  \end{center}
\end{figure}

\begin{figure}[htbp]
  \begin{center}
  \fbox{\epsfig{file=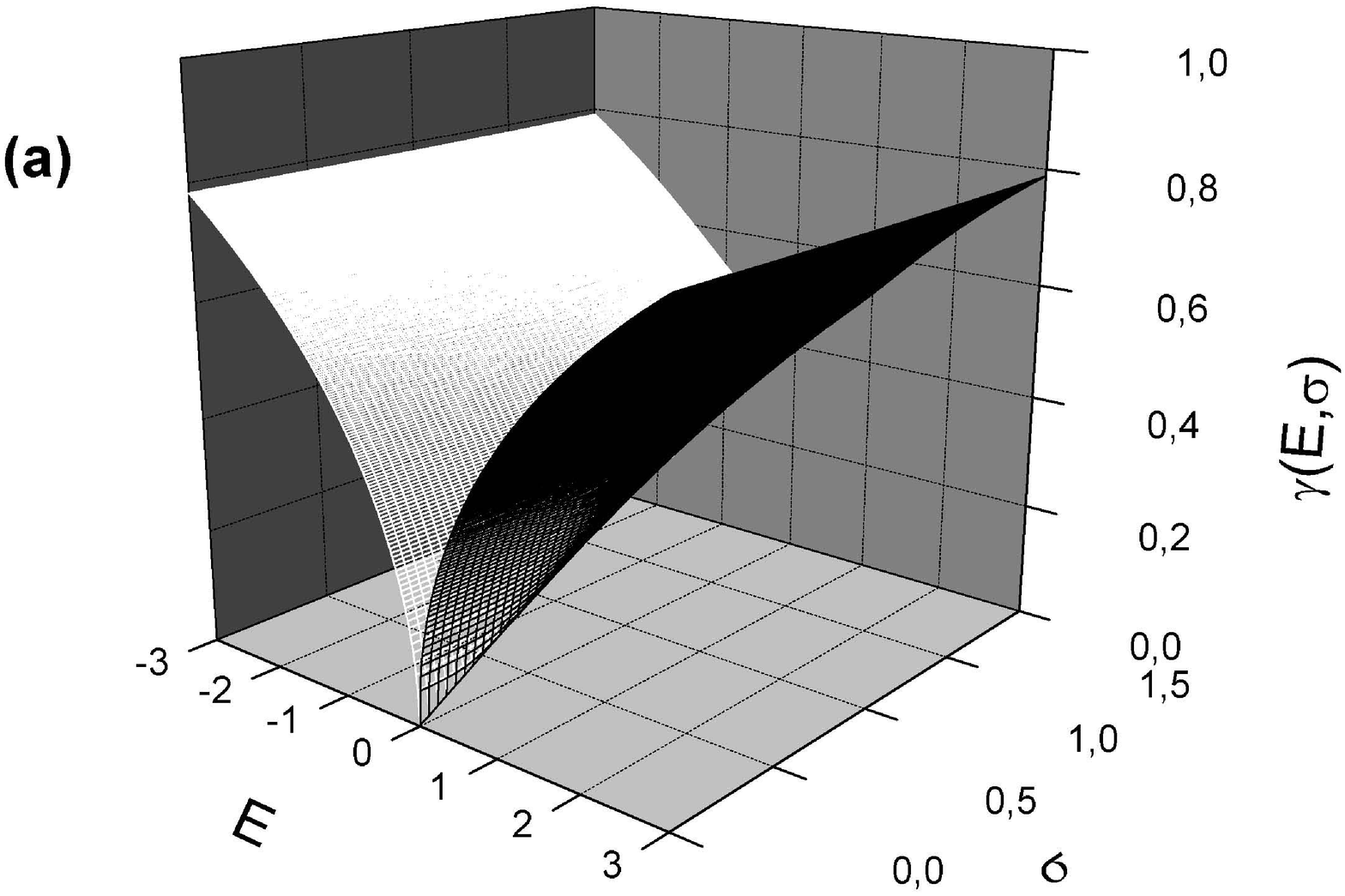,angle=0,width=4cm}}
    \fbox{\epsfig{file=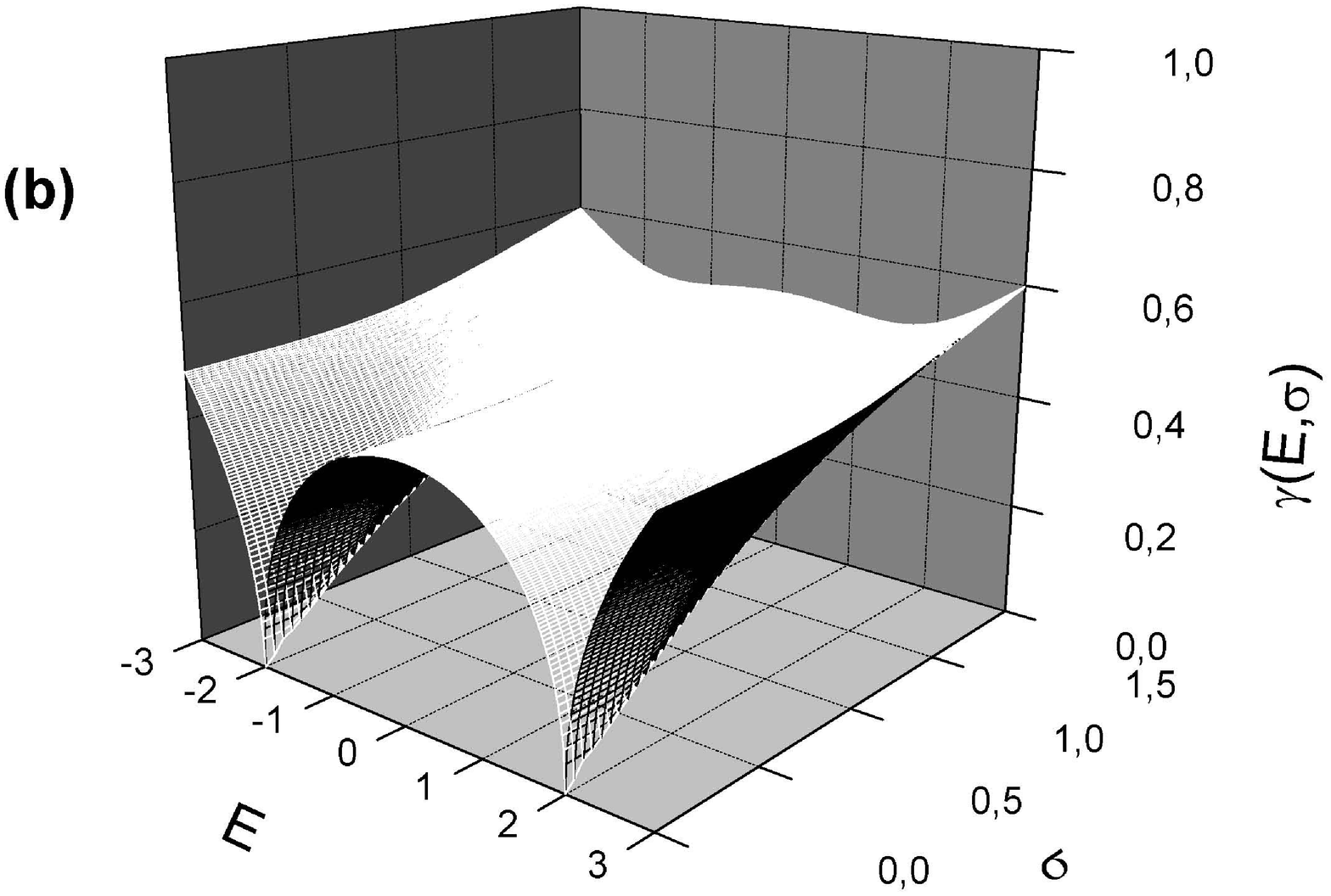,angle=0,width=4cm}}
    \fbox{\epsfig{file=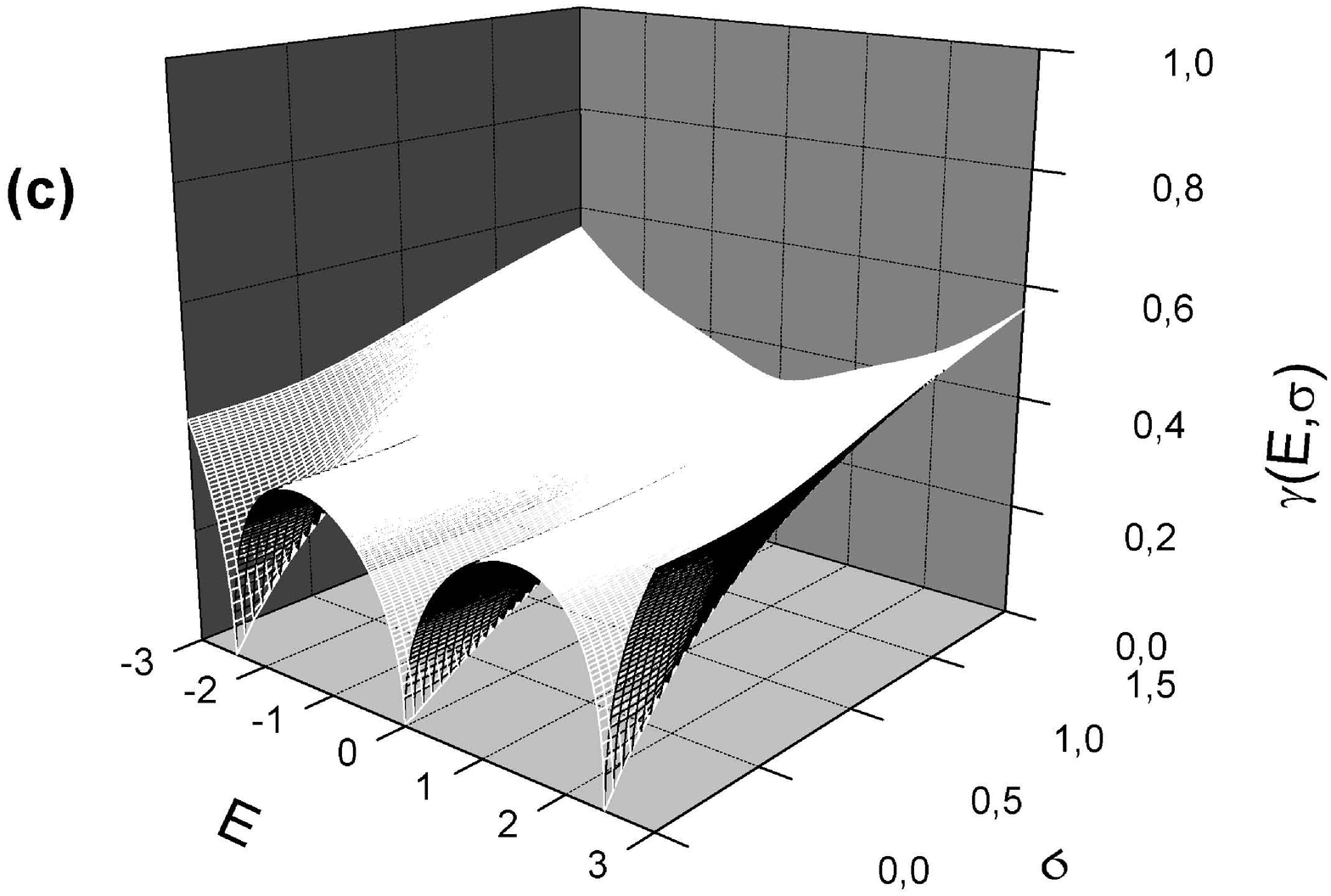,angle=0,width=4cm}}
    \caption{
    Surface plot $\gamma(E,\sigma)$ for $q=1$ (a), $q=2$ (b) and $q=3$ (c).
      }
    \label{fig: 2}
  \end{center}
\end{figure}

Therefore, in order to draw the phase diagram the analysis of the
stable filter $H_{-}(w)$, Fig.1, is sufficient. The
(\textit{unlimited}) existence region of the localized states
overlaps with that for the delocalized states (\textit{limited} in
both energy and disorder parameter). A number of such region
corresponds to a number $q$ of the Landau subbands, see Fig.1. Each
region is located near a center of old Landau subbands which existed
without disorder.

As a result, the phase diagram contains one-phase regions (only
localized states) and two-phase co-existence regions. The
delocalized state phase is defined by zero Lyapunov exponent
which needs no graphical illustrations. On the contrary,
the localized state phase is characterized by the function
$\gamma(E,\sigma)$ shown in Fig.2 in form of the surface plot. Note
also that the results in Fig.1 and Fig.2 correspond to the
\textit{thermodynamic limit}.

The generally-accepted point of view \cite{Huckestein95} is that in
the presence of a strong perpendicular magnetic field the extended
states appear in the centers of disordered-broadened Landau bands
and give rise to the integer quantum Hall effect. Instead, we
obtained a more complex situation. In each Landau subband there is a
region of phase co-existence. This region contains the critical
points ($E=E_c, \sigma=0$) where the localization length is
divergent. Moreover, the observed anomalies (the canyon bottom in
the surface plot of $\gamma(E,\sigma)$ in Fig.2) also lie in the
region of phase co-existence.

Since the traditional viewpoint is based mostly on the results of
numerical investigations, it is unclear how justified is here the
thermodynamic limit transformation. It was reasonably noted
\cite{Huckestein95} that we have to keep in mind that numerical
calculations always deal with finite systems, while, strictly
speaking, phase transitions exist only in infinite systems. It was
also believed there that finite-size scaling theory could serve as
theoretical framework to analyze data for finite systems, in order
to extrapolate results to infinite system size. Our analysis of the
phase diagram allows to understand better this conflicting
situation.

First of all, the general result should be mentioned. Both Fig.1 and
Fig.2 show that despite a quantitative difference, as a number of
the Landau subbands $q$ increases, the same qualitative picture
remains as it is observed for  $q=1$. There is the co-existence
region in each subband including the critical point $E_c$ and the
relevant anomaly in the Lyapunov exponent (localization length).
However, as $q$ increases, the region size systematically decreases.
One can expect that in systems with large $q$ the two-phase region
practically disappears. This is why there is some reasoning behind
the statement \cite{Markos06} that the model (\ref{Sch}) does not
exhibit a metal-insulator transition. This is true indeed but only
for a large number of Landau subbands.

The case of $q=1$ considered above allows us  to establish also a
link between systems with and without external magnetic field. As
was noted, under a strong magnetic field but for $p=q=1$,
Eq.(\ref{Sch}) coincides formally with that without magnetic field.
Therefore Fig.1a and Fig.2a illustrate also the phase diagram and
the $\gamma(E,\sigma)$-surface plot without magnetic field. In other
words, our correction of the statement \cite{Markos06,Huckestein95}
that in the presence of a strong  magnetic field extended states
appear in the centers of Landau subbands, namely, a demonstration
that the extended states could arise in fact in a wider region of
the phase-coexistence, sheds also light on theoretical
interpretation of the delocalized state existence without the
magnetic field. Our important conclusion is that an emergence of the
critical points is not a new effect caused by the magnetic field but
the fundamental feature of a system even without the imposed field.

In the main region denoted in Fig.1 as the ``localized states'' the
situation is simple and clear: a single phase exists for a given $E$
and $\sigma$ parameters. Since the analytical results correspond to
the average over the random potential realizations, the existence of
the delocalized states is here impossible. The random potential
realizations could give slightly different results but this means,
however, nothing but \textit{homophase} fluctuations. All quantities
are well defined and the terminology of the self-averaging
quantities is justified here. This region of the phase diagram is
not problematic also for standard numerical investigations. Indeed,
the numerical study deals with the fluctuations in a homophase
finite-size system, with the main focus on extrapolation of these
results with typically bad statistics to the thermodynamic limit.
Despite the fact that finite-size scaling is not strictly justified,
this faces no fundamental objections. The problem is that the
finite-size scaling works well in the cases when the region of the
phase diagram under study contains points with a divergent
correlation length. These points correspond to the relation
$\gamma(E_c,\sigma=0)=0$. However, as follows from Fig.2, such
points are absent in the homophase region. The more so, the values
of $\gamma(E,\sigma)$  are not small since they lie outside the
canyon bottom.

When considering the phase co-existence region in the phase diagram,
theory of the first-order phase transition is commonly used. It is
obvious that use of the two parameters, $E$ and $\sigma$, does not
define \textit{uniquely} the phase state of a system: an ensemble of
random potential $\varepsilon_{l,m}$ fluctuations  could be
\textit{heterophase}, not homophase ones. A study of such ensembles
faces a serious theoretical problem. It is common in experimental
studies to characterize a large sample in terms of a single phase,
giving no statistical analysis. However, in theoretical studies the
macroscopic features are associated with the ensemble average. It is
obvious that the averaging over the heterophase ensemble has no
physical sense: this is true only for characteristics of individual
homogeneous phases, self-averaging quantities lose here any sense
\cite{Kuzovkov02}.

The standard idea  that an increase of the sample length $L$ brings
us to the situation where the self-averaging effect automatically
takes care of the ensemble statistical fluctuations and the
statistical error can be estimated from sampling different
realizations of the disorder \cite{Huckestein95} does not work here.
An additional problem is that our approach
\cite{Kuzovkov02,Kuzovkov04,Kuzovkov06}  suggest no method to
estimate the weight (probability) of individual phases.

Therefore, the phase co-existence region contains the critical
points $(E=E_c,\sigma=0$) where the localization length is
divergent. However, the finite-size scaling cannot be used here,
since the situation differs qualitatively from that for the
second-order phase transition. That is, observation of the critical
points does not necessarily justify use of the finite-size scaling.

The very fact of the heterophase ensemble existence produces
fundamental difficulties for numerical investigations. In finite
systems the phase concept is not defined, thus this is
impossible to separated statistical contributions of the two phases.
Indeed, how could one distinguish large but rare homophase
fluctuation on a given parameter in a finite (typical small) size
and the heterogeneous fluctuation, especially when the statistics is
quite poor (relatively small number of random potential
realizations)? We are not able to suggest any new algorithm for
analysis of the relevant numerical investigations, we can only
discuss the consequences of the heterogeneous ensemble existence.

\subsection{Homophase interpretation}

The possibility of phase co-existence is neglected tradicionally in
numerical investigations \cite{Markos06}. It is believed that even
if the phase diagram consists of the two regions with two phases, in
each of these regions the relation between parameters $E$, $\sigma$
and the phase is unique. In other words, an existence of only
homophase ensembles is assumed, which could arise in the case of the
second-order phase transitions as well as without transitions. Let
us discuss what could be a result of the substitution of a
heterophase ensemble for homophase one, while performing numerical
investigations. We will demonstrate that the general statement
\cite{Huckestein95} follows uniquely from the homophase
interpretation.

1) In the phase co-existence region both realizations with
delocalized state, $\gamma\approx 0$, and localized state,
$\gamma>0$, exist. As was said above these contributions cannot be
separated (definition of a phase exists only for infinite systems
\cite{Huckestein95}), and the statistical analysis leads to a
formally calculated average $\langle \gamma \rangle>0$. This is
equivalent to the statement that the system does not exhibit a
metal-insulator transition (formally  $\langle \gamma \rangle>0$
means existence of the localized states only). Therefore the
homophase interpretation of 2-D numerical investigations with and
without magnetic field leads to well-known conclusions
\cite{Markos06,Huckestein95}

2) At the same time, the heterophase fluctuations should manifest
themselves in the statistics as \textit{abnormally} strong
fluctuations of the $\gamma$ parameter, which is unusual for
homophase systems and bring into question the treatment of the
$\gamma$ parameter as the self-averaging quantity. The situation is
complicated by the presence in the same region of the critical
points ($E=E_c,\sigma=0$) where the localization length diverges.
Since such critical points arise typically in the second-order
transitions, the quick conclusion suggests this interpretation with
strong homophase fluctuations nearly the critical points instead of
the real heterophase fluctuations. Moreveor, the observation of
critical points could be also used to justify the finite-size
scaling.

3) The anomaly of the Lyapunov exponent $\gamma(E,\sigma)$ near the
critical points is well pronounced in Fig.2. This specific anomaly
determined by the numerical methods, with a well pronounced
peculiarity of the canyon bottom in the plot $\gamma(E,\sigma)$
permits a traditional interpretation \cite{Huckestein95} about
critical points and the delocalized states.

The analysis above clearly demonstrates that the thermodynamic limit
has no correct solution by means of numerical investigation. The
extrapolation in terms of the finite-size scaling assumes an
existence of the homophase ensemble. However, there is no proof of
the the existence of such an ensemble within the same method. The
heterophase idea is not constructive since it suggests no
alternative to the finite-size scaling. The importance of exact
analytical solutions in such conflicting situations is self-evident.

\section{Conclusion}

We have shown that the contradiction between the results of the
present analytical approach (see also previous papers
\cite{Kuzovkov02,Kuzovkov04,Kuzovkov06,Kuzovkov07,Reply}) with those
of numerical investigations \cite{Markos06,Huckestein95} arises due
to different interpretations of the phase diagram. Our result
\cite{Kuzovkov02,Kuzovkov04,Kuzovkov06} correspond to the
thermodynamic limit where the phase concept is well defined
\cite{Baxter}. It was also shown that the Anderson localization
problem is characterized by the phase diagram with specific phase
co-existence regions which need analysis in terms of first-order
transitions. In those regions where the ensemble of heterophase
fluctuations occurs the idea of self-averaging quantities fails.
Analysis in the phase co-existence region is impossible for the
finite systems since the phase concept is not defined here.
Therefore, our analytical results
\cite{Kuzovkov02,Kuzovkov04,Kuzovkov06,Kuzovkov07,Reply} cannot be
reproduced by means of numerical investigations since we cannot
suggest any new algorithms for the numerical result analysis.

In its turn, the main numerical investigation results
\cite{Markos06,Huckestein95}  were obtained namely for the
finite-size systems. Transition to the thermodynamic limit is
traditionally performed in numerical investigations by means of the
finite-size scaling. The latter assumes (directly and indirectly)
existence of the ensemble of homophase fluctuations which permits
result extrapolation to the infinite single-phase system. This
approach fails in the case of two phase co-existence. Namely a wrong
use of the hypothesis of a single phase in the heterophase case
leads to incorrect theoretical conclusion about phase-transition
absence in 2-D systems which obviously contradicts experimental data
\cite{Abrahams01,Kravchenko04}. Simultaneously, even in the
framework of incorrect homophase interpretation numerical methods
are able to detect phase co-existence regions, interpreting these as
the critical points with relevant delocalized states.

\ack{This work was partially supported by grant  No.05.1704 of the Latvian
Council of Science.  Author is indebted to E. Kotomin and W. von
Niessen for detailed discussions of the paper.}


\end{document}